\documentstyle[preprint,aps,12pt,floats]{revtex}
\tightenlines
\newcommand{\beg}{\begin{equation}\label}
\newcommand{\en}{\end{equation}}

\newcommand{\bld}[1]{\mbox{\boldmath{$#1$}}}
\begin{document}
\thispagestyle{empty}
\begin{flushright} TTP 98-35\end{flushright}\vspace{1cm}

\begin{center}\begin{Large}\begin{bf}
Non-perturbative energy levels of two-fermion bound states\end{bf}\end{Large}\\
\vspace{1cm}\begin{Large}
Viktor Hund and Hartmut Pilkuhn\end{Large}\\ \begin{large}
 Institut f\"ur Theoretische Teilchenphysik,\\ Universit\"at,
D-76128 Karlsruhe, Germany\end{large}\\
(e-mails:vh \& hp@particle.physik.uni-karlsruhe.de)
\end{center}\vspace {1cm}
\begin{abstract}{For the S-states of positronium and muonium, the terms
    of an expansion of energy levels in powers of the fine structure
    constant $\alpha$ are also members of a ``recoil series''. The first
    two terms of that series are calculated to all orders in $\alpha$.}
\begin{center}{PACS number: 36.10.Dr,
    03.65.Pm}\end{center}\end{abstract}\vspace{0.8cm}

The calculation of energy levels to the order $\alpha^6$ for the
S-states of positronium and muonium $(e^-\mu^+)$ by Pachucki 
\cite{Pach1,Pach2,Pach3} has recently been
confirmed for equal masses by entirely analytic methods \cite{Czar}. 
Both calculations use non-relativistic quantum electrodynamics (NRQED) 
\cite{Cas}. 
The only two-fermion equation which is solved 
nonperturbatively is the Schr\"odinger equation for a particle of reduced 
mass $\mu=m_1m_2/m \; (m=m_1+m_2)$ in a Coulomb potential $V=-\alpha/r\;(\hbar
=c=1)$. Apart from a state-independent function $F(\mu/m)$
which is only numerically known for arbitrary $m_1/m_2$, 
all terms to order $\alpha^6$ and $\alpha^6
\log\alpha$ can also be arranged in a finite series in $\mu/m$ for the total 
cms energy $E$, 
\beg{1} E=m+E_1+E_2+E_{\rm log}+E_3,\en
where $E_i$ is of the order $\mu(\mu/m)^{i-1}$, and $E_{\rm log}$ contains 
logarithms of mass ratios. In the following, $E_1$ and $E_2$ are derived to all
orders in $\alpha$, and most of the sixth-order terms $E_3^{(6)}\sim \alpha^6
\mu^3/m^2$ of $E^{(6)}$
are confirmed (radiative corrections will be omitted). In combination with 
the complete expression for $E^{(6)}$, the series (\ref{1}) increases the 
precision of QED bound state calculations for systems with $m_1\ne m_2$. Our 
method also reduces the gap between relativistic and non-relativistic 
expansions.

The history of relativistic recoil expansions is long and sad
\cite{Eric,Sap,Eides}. Frequently, one starts from a Dirac equation
for an electron of mass $m_1$ in the potential $V(r)$. The subsequent
evaluation of recoil corrections of the order of $m_1/m_2$ shows that
most terms can be taken care of by replacing $m_1$ by $\mu$ in the Dirac
equation. Empirically then \cite{Eric}, $E_1$ follows from that
equation.
Similarly, the $\mu^2/m$-form has been found to order $\alpha^4$ for the
hyperfine splitting, which is part of $E_2$. But there has been no
indication that
to a given order of $\alpha$, the series (\ref{1}) would end after a finite
number of terms. 
For some P-states,
hyperfine mixing requires in fact infinitely many terms already at the  
order $\alpha^4$. The quantitative mixing disagrees with the standard
hyperfine operator of the Dirac equation. On the contrary, this mixing
follows easily from the Breit operators of NRQED. At the end, the Dirac
equation approach has been completely abandoned for two-body systems.
It is then remarkable that the nonperturbative use of the Schr\"odinger
equation leads to the form (\ref{1}), in which $E_1$ is given precisely
by the Dirac equation with reduced mass. For a state of angular momentum 
$j$, principal quantum number $n$ and with the abbreviation $j+\frac{1}{2}=
j_+$, the $\alpha^6$-part of $E_1$ is
\beg{2} E_1^{(6)}=-\mu\alpha^6\left(j_+^{-3}+3n^{-1}j_+^{-2}-6n^{-2}j_+^{-1}+
\frac{5}{2}n^{-3}\right)/8n^3,\en
from which Pachucki's result \cite{Pach2} follows for $j_+=1$.\\
We have recently derived a relativistic, Dirac-like two-fermion equation 
from perturbative QED \cite{ruth}, which explains this mystery and renders
the calculation of $E_2$ trivial, again to all orders in $\alpha$. The 
evaluation of $E_{\rm log}$
and of the state-independent $F(\mu/m)$ requires the calculation of loop 
integrals, which remains to be done.
The progress arises from the strict use of relativistic two-body kinematics 
in the first Born approximation, and from the reduction to $8\times 8$ 
components before Fourier transforming. 
Any system of two free particles of masses $m_1$ and $m_2$ satisfies the cms
equation $(p^2-k^2)\psi=0$, where the eigenvalue $k^2$ can be expressed in 
terms of $E^2, m_1^2$ and $m_2^2$. It can be brought into the form $k^2=
\varepsilon^2-\mu_E^2$,
\beg{3} \mu_E=m_1m_2/E, \quad \varepsilon= (E^2-m_1^2-m_2^2)/2E.\en
The free equation is converted into an explicit eigenvalue equation for $E^2$
by the substitution
\beg{4} {\bld\rho}=\mu_E{\bld r},\quad {\bld p}_\rho={\bld p}/\mu_E, \quad
\widehat\varepsilon=\varepsilon/\mu_E=(E^2-m_1^2-m_2^2)/2m_1m_2.\en
When the interaction is added, the Coulomb potential $V(r)$ is transformed 
into $V(\rho)$. The resulting dimensionless Dirac equation is \cite{ruth}
\beg{5} \left(\beta+\gamma_5({\bld\sigma_1}+{\bld\sigma_{hf}}){\bld p}_\rho +
V(\rho)-\widehat\varepsilon\right)\psi=0,\en
\beg{6} {\bld\sigma_{hf}}=-i{\bld\sigma_1}\times{\bld\sigma_2}V(\rho)
m_1m_2/E^2.\en
The ${\bld\sigma_1}$ and ${\bld\sigma_2}$ are Pauli matrices; the product 
$\gamma_5{\bld\sigma_1}$ is normally written as $\bld\alpha$. For comparison, 
the hyperfine ${\bld\sigma_{hf}}$ of atomic theory sets $E=m_2$ and replaces 
$V{\bld p}$ by $[V,{\bld p}]/2$.\\
To begin with, we evaluate the hyperfine energies $E_{hf}$ by first-order 
perturbation theory. For orbital angular momentum $l=j\pm \frac{1}{2}$ and 
total angular momentum $f=j\pm \frac{1}{2}$, they are
\beg{7} E_{hf}=2\frac{m_1^2m_2^2}{E^2}\alpha^4\frac{2(f-j)}{f+1/2}(j_+^2
\widehat\varepsilon-\kappa_D/2)/(4\gamma^3-\gamma),\en
\beg{8} \kappa_D=2(l-j)j_+, \qquad \gamma^2=j_+^2-\alpha^2.\en
Special cases of this formula are found in \cite{Rose}. To order $\alpha^6$
and for $\kappa_D=-j_+$, the quotient of the last two brackets in (\ref{7}) 
is
\beg{9}\frac{j_+^2\widehat\varepsilon-\kappa_D/2}{4\gamma^3-\gamma}=
\frac{1}{2jj_+}\left\{1+\alpha^2\left[\frac{1}{j(j+1)}+\frac{1}{2j_+^2}+
\frac{3}{2nj_+}-\frac{3}{2n^2}-\frac{j_+}{2n^2(j+1)}\right]\right\}.\en
The hyperfine splitting is defined as $\Delta E_{hf}=E(f=j+\frac{1}{2})-
E(f=j-\frac{1}{2})$.
For S-states, $l=0$ implies $j_+=1,\;\;2(f-j)/(f+\frac{1}{2})=\frac{2}{3}$ for 
$f=1$ and $-2$ for $f=0$, which makes a factor $\frac{8}{3}$ in 
$\Delta E_{hf}$. Pachucki's result for the part $E^{(6)}_{2, hf}$ of 
$E^{(6)}_{2}$ follows by approximating $E^2\approx m^2$ 
in the denominator of (\ref{7}). The remaining two terms of $E_2$ appear in 
the hyperfine-averaged shift $\bar E = \frac{3}{4}E(f=1)+\frac{1}{4}E(f=0)$.
They are conveniently evaluated by a non-relativistic reduction:\\
With the approximation $E^2\approx m^2$ in the hyperfine operator, (\ref{5}) 
is an explicit eigenvalue equation that permits the standard reduction by 
elimination of the small components. The resulting Schr\"odinger equation is 
\beg{10} (1+{\bld p}_\rho^2/2+V(\rho)-\widehat\varepsilon_{Sch})\psi_{Sch}=0.
\en
It has the familiar non-relativistic eigenvalues, $\widehat\varepsilon_{Sch}-1
=-\alpha^2/2n^2$.
The equation becomes quite powerful when its centrifugal barrier 
$l(l+1)/\rho^2$ is replaced by an effective barrier $l'(l'+1)/\rho^2$, which 
includes the lowest-order spin-orbit and hyperfine couplings:
\beg{11} l'-l\equiv \delta l=\alpha^2\left(-\frac{1}{2j_+}+\frac{2(f-j)}{f+
\frac{1}{2}}\frac{\mu}{m}\right).\en
The principal quantum number $n=n_r+l+1$ gets replaced by $n^*=n_r+l'+1=n+
\delta l$, and the eigenvalues $E^2_{Sch}$ follow from (\ref{10}) as
\beg{12} E^2_{Sch}-m^2=-\alpha^2m_1m_2/{n^*}^2\approx -\alpha^2m_1m_2 (1-
\delta l/n)^2/n^2.\en
To order $\alpha^6$, one obtains 
\beg{13} E_{Sch}-m=-\alpha^2(1-\delta l/n)^2\mu/2n^2-\alpha^4\mu^2 (1-
4\delta l/n)/8mn^4-\alpha^6\mu^3/16m^2n^6.\en
This expression contains both terms of $\bar E^{(6)}_2$ and two of the three 
terms in $\bar E^{(6)}_3$ \cite{Pach2} (note that $(\delta l)^2$ is quadratic 
in the hyperfine interaction). The third term arises from the S-D-mixing in 
second order perturbation theory and is not calculated here, $E_{SD}^{(6)}=-
\frac{4}{9}\delta_{f,1}\alpha^6\mu^3/m^2n^5$. It appears is $\bar E_3$ with a 
factor $\frac{3}{4}$, and in $\Delta E_{3, hf}$ with a factor 1. \\
There are two more $n^{-5}$-contributions to $\Delta E_{3, hf}$, one 
from the hyperfine part of $\delta l, \;\; \delta_{hf}=2\alpha^2\frac{\mu}{m}
(f-j)/(f+\frac{1}{2})$
in (\ref{13}) (in the bracket following $\alpha^4\mu^2$), and one from setting 
$E^2=m^2-\alpha^2m_1m_2/n^2$ in (\ref{7}), which effectively enlarges all 
hyperfine effects by a factor $1+\alpha^2\mu/mn^2$. The total coefficient of 
$\alpha^6\mu^3/m^2n^5$ is then $-\frac{4}{9}+\frac{4}{3}+\frac{8}{3}=
\frac{32}{9}$, in agreement with \cite{Pach1}.\\
For higher orders in $\alpha$, the expansion (\ref{1}) is conveniently 
replaced by the simpler expansion for $\widehat\varepsilon -1=
(E^2-m^2)/2m_1m_2$
\beg{13a} \widehat\varepsilon=1+\widehat\varepsilon_1+\widehat\varepsilon_2+
\widehat\varepsilon_{\rm log}+\widehat\varepsilon_3,\en
in which $\widehat\varepsilon_1$ is pure ``Dirac'' as in $E_1$, and 
$\widehat\varepsilon_2$ is pure ``hyperfine''. In other words, the 
non-hyperfine terms of $E_2$ are canceled to all orders in $\alpha$.\\
There is one more hyperfine contribution to $\Delta E_{3, hf}$ \cite{Pach1}
which does not follow from (\ref{5}):
\beg{14} \Delta E_{3, hf}=2\frac{\mu^3}{m^2}\,\alpha^6\,\frac{2(f-j)}{f+
\frac{1}{2}}\left(\ln{\frac{n}{\alpha}}+\frac{1}{n}-\frac{7}{6}-
\sum_{i=1}^{n-1}\frac{1}{i}\right).\en
It indicates that the combination of Dirac and hyperfine operators is still 
incomplete, at least for S-states. This is seen also for $m_1=m_2$, where 
the combination $\beta+V-\widehat\varepsilon$ of (\ref{5}) reduces to $V-
E^2/2m_1^2$ in the small components. In parapositronium, the same factor 
appears in the combination $({\bld\sigma_1}+{\bld\sigma_{hf}}){\bld p}$ in 
the large components and cancels out.
The resulting differential equation is reduced to the confluent hypergeometric 
one, in a shifted variable $r'=r-2\alpha/E$. Its singularity at $r'=0$ occurs 
at $r=2\alpha/E$, which excludes any bound state interpretation.\\
Further progress  in the calculation of $E_3$ and $E_{\rm log}$ in (\ref{1}) 
will require the recalculation of Feynman diagrams without kinematical 
approximations. The resulting formulas should not depend on the sign of $E$, 
as a consequence of $\cal CP$-invariance \cite{Malv}. In the integration over 
loop momenta, one normally expands the zero-components $p_i^0$ of the fermion 
momenta about their individual external values. To maintain 
$\cal CP$-invariance the product $\Sigma_1\Sigma_2$ of the fermion spin 
summations $\Sigma_i=p_i\gamma_i+m_i$ may be rearranged as follows:
\beg{16} \Sigma_1\Sigma_2=2p_1\gamma_1p_2\gamma_2+2m_1m_2-(p_1\gamma_1-m_1)
(p_2\gamma_2-m_2).\en
The last product contributes little; it vanishes on either mass shell, but its 
$\cal CP$-invariant evaluation would have to include negative values of $p^0_1$
and $p_2^0$. In the first term, on the other hand, one may keep $p^0_1p_2^0$ 
positive without breaking $\cal CP$.\\
The appearance of a factor $E^{-2}$ in the hyperfine operator (\ref{6}) has 
been confirmed here for the first time, by comparison with the $\alpha^6\mu^3/
m^2$-terms \cite{Pach1}. Outside QED, a similar effect seems to exist in the 
hyperfine splitting between vector $(1^-)$ and pseudoscalar $(0^-)$ mesons. 
The splitting in $E^2,\;\;\Delta=E^2(1^-)-E^2(0^-)$, increases uniformly from 
0.48 GeV$^2$ for the heavy $b$ quarkonium \cite{Rev} to 0.57 GeV$^2\approx 
m^2_\rho$ for the $\rho-\pi$ splitting \cite{Mannel}. If the light pseudoscalar
mesons could ever be treated by a potential model, their $E^2(0^-)$ would be 
lowered by the occurrence of $1/E^2$ in their hyperfine operator. The effect 
would become extreme in the limit $E^2\to 0$.

\medskip

One of the authors (H.P.) would like to thank Dr. J.O. Eeg for the kind 
hospitality extended to him at the physics institute of the University of 
Oslo, where this work was begun. \\
Helpful comments by Th. Mannel, K. Melnikov, and K. Pachucki are gratefully 
acknowledged. \\
This work has been supported by the Volkswagenstiftung.

\end{document}